\documentclass{article}
\usepackage{amssymb}

\newtheorem{theorem}{Theorem}
\newtheorem{acknowledgement}[theorem]{Acknowledgement}

\input{tcilatex}
\begin{document}

\title{Non-Euclidean Newtonian Cosmology}
\author{John D. Barrow \\
DAMTP, Centre for Mathematical Sciences,\\
University of Cambridge, Wilberforce Rd.,\\
Cambridge CB3 0WA, UK}
\maketitle

\begin{abstract}
We formulate and solve the problem of Newtonian cosmology under the
assumption that the absolute space of Newton is non-Euclidean. In
particular, we focus on the negatively-curved hyperbolic space, $H^{3}$. We
point out the inequivalence between the curvature term that arises in the
Friedmann equation in Newtonian cosmology in Euclidean space and the role of
curvature in the $H^{3}$ space. We find the generalisation of the
inverse-square law and the solutions of the Newtonian cosmology that follow
from it. We find the generalisations of the Euclidean Michell 'black hole'
in $H^{3}$ and show that it leads to different maximum force and area
results to those we have found in general relativity. We show how to add the
counterpart of the cosmological constant to the gravitational potential in $%
H^{3}$ and explore the solutions and asymptotes of the cosmological models
that result. We also discuss the problems of introducing compact topologies
in Newtonian cosmologies with non-negative spatial curvature.
\end{abstract}

\section{Introduction}

When Newton formulated his theory of gravity he assumed time to be linear,
with a rate of flow that was unchangeable. He assumed space to the absolute,
unchanging, and Euclidean: 'the divine sensorium' \cite{newt}. All motion
was played out on this Euclidean space without influencing the geometry of
space or the rate of flow of time. Newton's assumption of 3d Euclidean
geometry, $E^{3}$, was tied to the belief that had existed for centuries
that Euclidean geometry was part of the ultimate truth about the nature of
the universe. It was not just a mathematical model. If a philosopher or
theologian was challenged that their quest for ultimate truth was doomed to
fail because the human mind was too limited, then they could just point to
Euclid's geometry as an example of our success in finding a part of the
ultimate truth \cite{pi}.

Yet we can see that there was no need to formulate Newton's theory of
gravity on Euclidean space. In this paper we are going to look at what
happens to cosmology when Newtonian gravity is formulated on Lobachevsky's
space of constant negative curvature -- ie hyperbolic 3-space, $H^{3}$, in
modern terminology. The situation in a positively curved space of constant
curvature will also be briefly discussed.

The subject of Newtonian cosmology has a long history, beginning with Milne
and McCrea's deduction of the counterpart to the Friedmann universes \cite%
{MMc}. The general relationship between Newtonian and relativistic gravity
has been partly obscured because most of the studies of the former have been
carried out in the context of homogeneous and isotropic universes where the
incompleteness of Newtonian gravity is hidden. When anisotropies are present
the lack of any propagation equations for the shear tensor components is
notable \cite{schuck}. As a result, the rigorous mathematical studies of the
asymptotic behaviour of the unbound Newtonian N-body problem obtain results
for the radius of gyration (or moment of inertia) of the system but give no
information about the shape anisotropy \cite{saar}.

One of the features we will highlight is the difference between Newtonian
cosmology on an absolute curved space of constant curvature and the
Newtonian version of the general relativistic Friedmann universes where the
metric curvature constant $k$ no longer describes any curvature because
Newtonian cosmology in that Milne-McCrea context is always Euclidean and the
constant labelled by $k$ arises from the conservation of the total energy of
motion. The possibility of non-Euclidean Newtonian cosmologies is ignored in
all textbook expositions of general relativistic cosmology that use the
Newtonian formulation of cosmology to motivate it.

Newton's inverse-square law in $E^{3}$ has a form that is inversely
proportional to the area of a 3d ball of the required curvature: $4\pi r^{2}$%
. So, in $E^{3}$, we have the familiar force law, $F(r),$ and gravitational
potential, $\Phi (r)$

\begin{equation}
F\propto \frac{1}{r^{2}},\text{ \ }\Phi \propto \frac{1}{r}.  \label{1}
\end{equation}

\bigskip Similarly, in $H^{3}$, using the area element, $4\pi R^{2}\sinh
^{2}(r/R)$, we have

\begin{equation}
F\propto \frac{1}{R^{2}\sinh ^{2}(r/R)},\text{ \ \ }\Phi \propto \coth (r/R),
\label{2}
\end{equation}%
where the constant $R$ is the curvature radius. In the Euclidean space limit 
$R\rightarrow \infty ,$the formulae eq.\ref{2} reduce as expected to eq.\ref%
{1} \ \footnotetext{%
If you want $\Phi (r)\rightarrow 0$ as $r\rightarrow \infty $ then the
constant of integration in the Poisson equation can be used to produce a
solution $\Phi (r)=-\frac{GM}{r}\coth (r/R)+\frac{GM}{R}$ with that
behaviour \ but we shall not use that form.}. This expression for the area
element was first given by Lobachevsky in 1835 in the form $\pi
(e^{r}-e^{-r})^{2}$ in \ ref. \cite{lob}. The connection with inverse-square
laws of physics was also appreciated by Bolyai after 1832 (but surprisingly
not by Gauss) and the analytic form for the Newtonian potential in $H^{3}$
was given by Schering in 1870 \cite{scher}. One of the motivations for these
deviations from the $E^{3}$ inverse-square law of Newton was the desire to
avoid divergence of the total mass, gravitational potential, gravitational
force and tidal force at infinity. This was possible by changing the
inverse-square law of Newton or by assuming space was not $E^{3}$ \cite{nort}%
. The history of these developments and the application of non-Euclidean
geometry to Bertrand's problem for potentials that give circular orbits for
the Kepler problem, Newton's spherical property, and the force laws given
above is rather confused with incorrect attributions to first studies; a
good overview of the history is given in refs. \cite{shch, diac} and we will
not repeat it here. The case of a positively curved space of constant
curvature, $S^{3}$, is a simple complex transform that changes the $\sinh $
to $\sin $, so yields

\begin{equation}
F\propto \frac{1}{R^{2}\sin ^{2}(r/R)},\text{ \ \ }\Phi \propto \cot (r/R).
\label{3}
\end{equation}

\section{Solutions of the Poisson Equation in $H^{3}$}

We formulate Newtonian cosmology on a fixed non-Euclidean space by solving
Poisson's equation

\begin{equation}
\nabla ^{2}\Phi =4\pi G\rho ,  \label{4}
\end{equation}%
with $\nabla ^{2}$ defined on a negatively-curved space in ($r,\theta ,\phi $%
) coordinates by

\begin{equation}
\nabla ^{2}=\frac{1}{r^{2}}\frac{\partial }{\partial \rho }r^{2}\frac{%
\partial }{\partial \rho }+\frac{1}{r^{2}}\left( \frac{1}{\sin \theta }\frac{%
\partial }{\partial \theta }\sin \theta \frac{\partial }{\partial \theta }+%
\frac{1}{\sin ^{2}\theta }\frac{\partial ^{2}}{\partial \phi ^{2}}\right) .
\label{4a}
\end{equation}

\bigskip

Specialising to $\Phi (r)$, the solution to $\nabla ^{2}\Phi =0$ or $\nabla
^{2}\Phi =$ $4\pi G\delta (x)\delta (y)\delta (z)$ for the gravitational
action on a test mass, $M,$where the arguments of the delta functions are
given by

\begin{equation}
x=r\sin \theta \cos \phi ,\text{ \ }y=r\sin \theta \sin \phi ,\text{ \ }%
z=r\cos \theta ,  \label{5}
\end{equation}

is%
\begin{equation}
\Phi (r)=-\frac{GM\coth (r/R)}{R},\text{ \ }F(r)=-\frac{\partial \Phi }{%
\partial r}=-\frac{GM}{R^{2}\sinh ^{2}(r/R)}  \label{6}
\end{equation}

to obtain the generalisation of Newton's inverse-square law for the
acceleration at a distance $r$ from a mass $M.$\footnote{%
To repeat the calculation for spherical space change $sinh$ to $sin$). In N
space dimensions, $H^{N}$, the required solution of Laplace's equation is $%
\Phi (r)\varpropto \coth ^{n-2}(r/R)$ for $n>2$ and $ln[tanh(r/R)]$ when $%
n=2.$}. This approaches the Euclidean results $\Phi =-GM/r$ and $F=-GM/r^{2}$
when $R>>r,$. We can use this equation to obtain the generalised Friedmann
equation for a Newtonian universe of constant negative curvature:

\begin{equation}
\ddot{r}=-\frac{GM}{R^{2}\sinh ^{2}(r/R)},  \label{7}
\end{equation}%
and integrate this equation to obtain the generalised Friedmann equation for
a Newtonian universe of constant negative curvature containing matter with
zero pressure (the cases of non-zero pressure could be investigated by
changing $\rho \rightarrow \rho +3p$ in Poison's eqn.):

\begin{equation}
\frac{\dot{r}^{2}}{r^{2}}=\frac{2GM}{Rr^{2}}\coth (\frac{r}{R})-\frac{K}{%
r^{2}},  \label{8}
\end{equation}%
with $K$ an integration constant. Note that this is not the same as the
general-relativistic Friedmann equation for a cosmology with a space of
constant negative curvature (although it reduces to it in the Euclidean $r<<R
$ limit).

\bigskip From eq. \ref{8} with $K=0$ we can find the scale factor of an
expanding universe $r(t)$ from

\begin{equation}
\int \sqrt{\tanh (r/R)}dr=\sqrt{\frac{GM}{R}}(t+t_{0}),  \label{9}
\end{equation}%
hence

\begin{equation}
\tanh ^{-1}(\sqrt{\tanh (r/R)})-\tan ^{-1}(\sqrt{\tanh (r/R)})=\sqrt{\frac{GM%
}{R}}(t+t_{0}).  \label{10}
\end{equation}%
We have $t_{0}=0$ if we set $r(0)=0$ and there is an initial singularity in
the density $\rho \propto r^{-3}\rightarrow \infty $ as $r\rightarrow 0$.

The case with $K\neq 0$ does not allow eq. (\ref{8}) to be integrated
exactly. In the $r\rightarrow \infty $ limit we have with $K=-1$:

\begin{equation}
\dot{r}^{2}\rightarrow \frac{2GM}{R}\{1+2\exp [-2r/R]\}-K,  \label{11}
\end{equation}%
and to leading order we have

\begin{equation}
\frac{1}{2}\dot{r}^{2}\rightarrow \frac{GM}{R}-K,  \label{12}
\end{equation}%
so for $K=-1$

\begin{equation}
r\propto t.  \label{13}
\end{equation}%
and the solution approaches the Milne universe, as in general relativity and
Euclidean Newtonian gravity.

\section{Non-Euclidean Black Holes}

It is possible to follow the Newtonian logic of Michell \cite{mitch} and
Laplace \cite{lap} (see ref \cite{mont} for a detailed history) in the case
of Newtonian gravity on $H^{3}$. When $K=0$ we can create a Newtonian 'black
hole' by seeking the dimensions of a spherical gravitating object with
escape velocity equal to that of light: $\dot{r}^{2}=c^{2}=2GMR^{-1}\coth
(r/R)$. Its radius is

\begin{equation}
r_{bh}=R\tanh ^{-1}(\frac{2GM}{Rc^{2}})=\frac{R}{2}\ln \left( \frac{1+\frac{%
2GM}{Rc^{2}}}{1-\frac{2GM}{Rc^{2}}}\right) ,  \label{14}
\end{equation}%
for the principal branch $-1<\frac{2GM}{Rc^{2}}<1$ and $r_{bh}\rightarrow
2GM/c^{2}$ in the Euclidean limit $R\rightarrow \infty $. As in Michell's
Newtonian black hole, the radius $r_{bh}$ is not an event horizon of no
return as in general relativity. It is a surface that requires an escape
velocity of $c$ in order to escape all the way to infinity. Any finite
distance from the black hole can be reached with a suitable launch speed.

If the surface area of the 'black hole' is $4\pi R^{2}\sinh ^{2}(r_{bh}/R)$
then this might still be a measure of entropy, as in general relativity:

\begin{eqnarray}
A &=&4\pi R^{2}\sinh ^{2}(r_{bh}/R)=\frac{4\pi R^{2}}{\coth ^{2}(r_{bh}/R)-1}%
=\frac{4\pi R^{2}}{\frac{R^{2}c^{4}}{4G^{2}M^{2}}-1}=...  \label{15} \\
&=&\frac{16\pi G^{2}M^{2}R^{2}}{R^{2}c^{4}-4G^{2}M^{2}}=\frac{4\pi R_{s}^{2}%
}{1-R_{s}^{2}/R^{2}}\simeq 4\pi R_{s}^{2}(1+R_{s}^{2}/R^{2}+..)>A_{s}, 
\nonumber
\end{eqnarray}%
where $R_{s}=2GM/c^{2}$ is the usual Schwarzschild radius and $A_{s}=4\pi
R_{s}^{2}$.

To next order $\tanh ^{-1}x=x+x^{3}/3+..$

\begin{equation}
r=R_{S}\left[ 1+\frac{R_{S}^{2}}{3R^{2}}+..\right] ,  \label{16}
\end{equation}%
where $R_{S}<R$ and

\begin{equation}
R_{S}=\frac{2GM}{c^{2}}.  \label{17}
\end{equation}

\section{Maximum Force}

Barrow and Gibbons \cite{BG2} have shown that in general relativity there is
strong evidence that there exists a maximum force, equal to $c^{4}/4G$ in
three space dimensions (but not in higher dimensions). This is the Planck
unit of force but Planck's constant does not appear and so it has
fundamental classical significance, like the Planck unit of velocity, $c$,
and the magnetic moment to angular momentum ratio \cite{BG1}. It is
interesting to ask what becomes of this bound in $H^{3}$. A simple guide is
to use the fact that when two Schwarzschild black holes touch horizons, the
force between their centres gives the maximum force\footnote{%
It is also the maximum cosmic string tension that arises when the deficit
angle caused by the string equals $2\pi .$}. Using two $H^{3}$ Newtonian
black holes of mass $M$ described by eq. \ref{14}, the maximum force between
them at separation $r_{bh}$ is

\begin{equation}
F=\frac{GM^{2}}{R^{2}\sinh ^{2}\left[ \tanh ^{-1}(2GM/Rc^{2}\right] }.
\label{17a}
\end{equation}

As $R\rightarrow \infty $, we recover the general relativity limit \cite{BG2}%
, as $F\rightarrow \frac{c^{4}}{4G}$. Keeping higher-order corrections as $%
R\rightarrow \infty $, we have

\begin{equation}
F\rightarrow \frac{GM^{2}}{R^{2}\sinh ^{2}\left[ \frac{2GM}{Rc^{2}}\left( 1+%
\frac{4G^{2}M^{2}}{3R^{2}c^{4}}\right) \right] },  \label{17b}
\end{equation}%
and so, to second order,

\begin{equation}
F\rightarrow \frac{c^{4}}{4G}\left[ 1-\frac{8G^{2}M^{2}}{3R^{2}c^{4}}\right]
<\frac{c^{4}}{4G},  \label{17c}
\end{equation}

and the maximum force goes to zero when $R=\sqrt{\frac{2}{3}}R_{s}$.

From eq. \ref{17a}, we see this reduction in the maximum force holds in
general because $\tanh ^{-1}(x)>x$ on $0<x<1.$Hence, we see that the
candidate maximum force is reduced from its value in the $E^{3}$ Newtonian
and general relativistic cases examined in ref. \cite{BG2}.

\section{Adding a "Cosmological Constant"}

The addition of the cosmological constant in the Euclidean Newtonian problem
amounts to the determination of the two forms of the gravitational potential 
$\Phi (r)$ that obey the Newton spherical property and make the external
gravity field of a sphere equal to that of a point mass at its centre of
equal mass. There are two potentials of this sort

\begin{equation}
\Phi \varpropto \frac{1}{r}\text{ \ \ and }\Phi \varpropto \Lambda r^{2},
\label{18}
\end{equation}

Both of these solutions were known to Newton and appear in the \textit{%
Principia, }but it was Laplace who showed that the general solution for the
spherical property for a potential in $E^{3}$ is the linear combination of
these two potentials, yielding \cite{BT},%
\begin{equation}
\Phi =\frac{-GM}{r}\text{ \ }+\text{ }\frac{\Lambda r^{2}}{6}.  \label{19}
\end{equation}%
In \ order to find the second potential in $H^{3\text{ }}$that is the
counterpart of the $\Phi \varpropto \Lambda r^{2}$ in $E^{3}$, we can use
the elegant analysis of the general Bertrand problem for closed circular
orbits in static spherically symmetric made by Perlick \cite{per} and
elaborated by Ballesteros et al \cite{ball}. Writing the spatial metric for
a general static spherically symmetric metric as

\begin{equation}
dl^{2}=h(r)^{2}dr^{2}+r^{2}(d\theta ^{2}+\sin ^{2}\theta d\phi ^{2}),
\label{20}
\end{equation}%
the radial Laplacian operator for the radial potential is then

\begin{equation}
\nabla ^{2}\Phi (r)=\frac{1}{r^{2}h(r)}\frac{d}{dr}\left( \frac{r^{2}}{h(r)}%
\frac{d\Phi }{dr}\right) .  \label{21}
\end{equation}

Hence, the solution of $\nabla ^{2}\Phi (r)=0$ is

\begin{equation}
\Phi (r)=\tint\limits^{r}\frac{h(r)}{r^{2}}dr.  \label{22}
\end{equation}%
The counterparts of eq. \ref{18} in general are again the inverse squares of
each other, as in $E^{3}$\cite{ball}:

\begin{equation}
\Phi _{1}(r)=A_{1}\left( \tint\limits^{r}r^{-2}h(r)dr+B_{1}\right) ,
\label{23a}
\end{equation}

\begin{equation}
\Phi _{2}(r)=A_{2}\left( \tint\limits^{r}r^{-2}h(r)dr+B_{2}\right) ^{-2},
\label{23b}
\end{equation}

\bigskip where $A_{1},A_{2},B_{1}$ and $B_{2}$ are constants. For the space
of constant curvature, $K$, familiar from general relativistic cosmology, we
have

\begin{equation}
\Phi _{1}(r)=\sqrt{r^{-2}-K},\text{ \ \ \ \ \ }\Phi _{2}(r)=\left(
r^{-2}-K\right) ^{-1}.  \label{24}
\end{equation}%
If we introduce the distance function to $r=0$ on the curved space, $\rho
_{K}=r=R\tanh (r/R)$, then we have for $K=-1,$noting that

\[
K=\frac{-1}{R^{2}},
\]

\bigskip we obtain the inverse-square pair \cite{per, ball, voz} \footnote{%
For the positively curved space ($K=+1$) the same results hold with the
transforms $\coth \rightarrow \cot $ and $\tanh \rightarrow \tan .$},%
\begin{eqnarray}
\Phi _{1} &=&-\frac{GM}{R}\coth \ \left( \rho _{K}/R\right) ,  \label{25a} \\
\Phi _{2} &=&\frac{\Lambda R^{2}}{6}\tanh ^{2}(\rho _{K}/R),\   \label{25b}
\end{eqnarray}

where the constant of proportionality has been labelled $\Lambda /6$ so as
to recover the $E^{3}$ proportionality constant in \ eq. \ref{19} in the
Euclidean limit $R\rightarrow \infty $, where $\Phi _{1}\propto -\frac{1}{%
\varrho _{K}}$ and $\Phi _{2}\propto \varrho _{K}^{2}.$ Thus, $\Phi _{2}$
gives us the $H^{3}$ equivalent of the $\Lambda $-term potential giving the
Newtonian spherical property and Bertrand's circular orbits like Newton's
potentials in eq. \ref{18} in $E^{3}$.

\bigskip What happens to the cosmological eqns. now? The force law (using
the $r$ coordinate now) is

\begin{equation}
F=-\nabla \Phi =\frac{-GM}{R^{2}\sinh ^{2}(r/R)}+\frac{\Lambda R}{3\cosh
^{2}(r/R)}\tanh (r/R).  \label{26}
\end{equation}%
At small $r$ , we recover the $E^{3}$ Newtonian \ result for the force on a
test particle $m$

\begin{equation}
F\rightarrow \frac{-GMm}{r^{2}\ }+\frac{m\Lambda r}{3\ }.  \label{27}
\end{equation}

There is a counterpart to the Newtonian de Sitter solution if we keep only
the second potential ($\Phi _{2}$) on the right-hand side of eq. \ref{26}

\begin{equation}
\ddot{r}=\frac{\Lambda R}{3\cosh ^{2}(r/R)}\tanh (r/R).  \label{28}
\end{equation}%
So at large $r>>R,$

\begin{equation}
\ddot{r}\rightarrow \frac{\Lambda r}{3\ },  \label{29}
\end{equation}

and we recover the de Sitter asymptote, $r(t)=\exp [t\sqrt{\Lambda /3}].$

In general for this case, we have

\begin{equation}
\dot{r}^{2}=\frac{2\Lambda R}{3}\int \frac{\tanh (r/R)}{\cosh ^{2}(r/R)}dr=%
\frac{2\Lambda R^{2}}{3}\tanh ^{2}(r/R)+C,  \label{30}
\end{equation}%
where $C$ is an integration constant. The solution for $C=0$ is

\begin{equation}
\sinh (r/R)=\exp [\sqrt{\frac{2\Lambda }{3}}(t+t_{0})]  \label{31}
\end{equation}%
where $t_{0}$ is an inessential constant. Thus, as $t\rightarrow \infty $,
we have The Milne asymptote

\begin{equation}
r\varpropto t  \label{32}
\end{equation}

This is understandable because the force goes to zero exponentially as $%
r\rightarrow \infty $: the universe looks like a vacuum solution.

If we keep both potential terms in eq. \ref{26}, then

\begin{equation}
\frac{\dot{r}^{2}}{r^{2}}=\frac{2GM}{Rr^{2}}\coth (r/R)+\frac{2\Lambda R^{2}%
}{3r^{2}}\tanh ^{2}(r/R)+\frac{C}{r^{2}},  \label{33}
\end{equation}%
and as $r\rightarrow \infty $, the first ('matter') term on the right-hand
side (from $\Phi _{1}$) is dominated by the curvature-like term $C/r^{2}$and
the $\Lambda /r^{2}$ terms. We will not unpick the various cases determined
by the size and sign of the ratio $\frac{2\Lambda R^{2}}{3C}$ but it is a
straightforward exercise$.$

\section{ Compact topologies}

In \ general relativistic cosmologies we are familiar with the possibility
that universes with zero or negative spatial curvature might have finite
volume due to a compact topology \cite{fried1922, fried1924, ellis}. The
simplest example would be a 3-torus topology for the zero-curvature
Friedmann universe. When universes become anisotropic the problem of
acceptable topologies becomes much more complicated and there are
anisotropic Bianchi type universes where compactification of the topology
forces the dynamics to be isotropic \cite{ash, BK}. However, unlike in
general relativity, the possibility of compactification of a negatively
curved or flat space in Newtonian cosmology seems to be prohibited. If we
integrate Poisson's eqn. over the 3-sphere

\begin{equation}
\frac{1}{4\pi G}\int_{V}\rho dV=\int_{V}\nabla ^{2}\Phi dV=\int_{V}\func{div}%
\func{grad}\Phi dV=\int_{S}\func{grad}\Phi dS=0  \label{34}
\end{equation}

The last two steps use the divergence theorem and the fact that the last
integral is over the boundary $S$ of the three-sphere, and  so is integral
is zero when the cosmological constant vanishes since the three-sphere has
no boundary. However, this implies that $\rho =0$ everywhere since $\rho
\geq 0.$ If we had included pressure then we would have arrived at the
conclusion that $\rho +3p=0,$\cite{chav, BT, FJT}. This argument can be
applied to our non-Euclidean cosmological model derived from eq. \ref{4} to
exclude compact spatial topologies in the absence of a cosmological constant.

\section{Conclusions}

We have formulated the theory of Newtonian gravity on a non-Euclidean
absolute space of constant curvature and investigated the case of
negatively-curved, 3-d hyperbolic space, $H^{3},$ in detail. We have also
indicated how the case of positively curved space can be obtained from it.
We derived the generalisation of Newton's inverse-square law in $H^{3}$ and
used it to develop a non-Euclidean Newtonian cosmology of the sort first
created by Milne and McCrea \cite{MMc} for Euclidean space and found
solutions of the generalised Friedmann equation. We showed how the curvature
term of general relativity that shows up in the Newtonian counterpart is
different to the terms governing the curvature of space in the $H^{3}$
Newtonian cosmology. We found the structure of Newtonian 'black holes' in $%
H^{3}$ by generalising the arguments first employed by Michell. We also
found the modifications to the maximum force condition in general
relativistic black holes using the $H^{3}$ 'black hole', showing that the
maximum force is reduced below its general relativistic value argued for
earlier by Barrow and Gibbons \cite{BG2}, while the area of the 'black hole'
is increased over its general relativistic value. Using the analysis of
Perlick \cite{per}, we then showed how to add the equivalent of a
cosmological constant to non-Euclidean gravity theory by using the Bertrand
criteria for closed circular orbits and Newton's spherical property to
generalise the sum of $r^{-1}$ and $r^{2}$ potentials first found by Laplace
to the inverse-square pair $\coth r$ and $tanh^{2}r$. We studied the
cosmological solutions and asymptotic forms that they display at small and
large times when this generalised cosmological constant term is present.
Finally, we make some remarks about the impossibility of compact topologies
in Newtonian gravity when space is flat or negatively curved.

\begin{acknowledgement}
\bigskip G.W. Gibbons is thanked for discussions and the STFC of the United
Kingdom for support.
\end{acknowledgement}

\end{document}